\documentclass[12pt,preprint]{aastex}
\slugcomment{}

\shorttitle{FALLBACK DISK AROUND 4U 0142+61}
\shortauthors{ERTAN ET AL.}

\begin{document}

\title{THE ANOMALOUS X-RAY PULSAR 4U 0142+61: A NEUTRON STAR WITH A GASEOUS FALLBACK DISK}
\author{\"{U}. ERTAN\altaffilmark{1}, M. H. ERKUT\altaffilmark{1},
K. Y. EK\c{S}\.{I}\altaffilmark{2}, AND M. A. ALPAR\altaffilmark{1}}

\affil{\altaffilmark{1}Sabanc\i\ University, 34956, Orhanl\i\, Tuzla, \.Istanbul, Turkey}
\affil{\altaffilmark{2}\.Istanbul Technical University (\.{I}T\"{U}), \.Istanbul, Turkey}


\begin{abstract}
The recent detection of the anomalous X-ray pulsar (AXP) 4U 0142+61
in the mid-infrared with the {\it Spitzer} observatory by Z.Wang and coworkers 
constitutes the first instance of a
disk around an AXP. We show, by analyzing earlier optical and
near-IR data together with the recent data, that the overall broadband data set  can be reproduced by a single model of an irradiated and viscously heated 
disk. 
\end{abstract}

\def\la{\raise.5ex\hbox{$<$}\kern-.8em\lower 1mm\hbox{$\sim$}}
\def\ga{\raise.5ex\hbox{$>$}\kern-.8em\lower 1mm\hbox{$\sim$}}
\def\be{\begin{equation}}
\def\ee{\end{equation}}
\def\ba{\begin{eqnarray}}
\def\ea{\end{eqnarray}}
\def\be{\begin{equation}}
\def\ee{\end{equation}}
\def\ba{\begin{eqnarray}}
\def\ea{\end{eqnarray}}
\def\m{\mathrm}
\def\d{\partial}
\def\R{\right}
\def\L{\left}
\def\a{\alpha}
\def\Mdot*{\dot{M}_*}
\def\Mdotin{\dot{M}_{\mathrm{in}}}
\def\Mdot{\dot{M}}
\def\Lin{L_{\mathrm{in}}}
\def\Rin{R_{\mathrm{in}}}
\def\rin{r_{\mathrm{in}}}
\def\Rout{R_{\mathrm{out}}}
\def\Rout{R_{\mathrm{out}}}
\def\Ldisk{L_{\mathrm{disk}}}
\def\Lx{L_{\mathrm{x}}}
\def\dEb{\delta E_{\mathrm{burst}}}
\def\dEx{\delta E_{\mathrm{x}}}
\def\Bb{\beta_{\mathrm{b}}}
\def\Be{\beta_{\mathrm{e}}}
\def\Rc{\R_{\mathrm{c}}}
\def\dMin{\delta M_{\mathrm{in}}}
\def\dM*{\delta M_*}
\def\Teff{T_{\mathrm{eff}}}
\def\Tirr{T_{\mathrm{irr}}}
\def\Firr{F_{\mathrm{irr}}}
\def\Av{A_{\mathrm{V}}}
\def\p{\propto}
\def\m{\mathrm}

\section{INTRODUCTION}

The zoo of young neutron stars contains a number of categories recognized by the distinct properties
of such stars discovered or identified in the last decade. The existence of these different types of neutron stars
strongly suggests that not all neutron stars are born with the same initial conditions,
nor do they follow the same evolutionary path as the familiar radio pulsars, of which the typical young example is usually
taken to be the Crab pulsar.
Anomalous X-ray pulsars (AXPs) and soft gamma-ray repeaters (SGRs)
\citep[see][for a review of AXPs and SGRs]{WT05} are widely accepted to be magnetars \citep{DT92}.
The "radio-quiet" neutron stars (also known as central compact objects [CCOs]; (Pavlov et al. 2004) and
dim isolated (thermal) neutron stars \citep[DTNs or DINs;][]{haberl05}
are probably related classes, although the latter are relatively older. Measured rotation periods of AXPs, SGRs and DINs
cluster in the narrow range of $3-12$ s. \citet{Alpar01} proposed
that the presence or absence, and properties, of bound matter with angular momentum
that is, a fallback disk,  may be among the initial
parameters of newborn neutron stars, in addition to magnetic dipole
moment and initial rotation rate. According to this scenario, the differences between isolated
pulsars, AXPs, SGRs, and DINs, as well as the CCOs in certain supernova remnants, are due to
different initial conditions, including the presence or absence and properties of fallback disks.

The suggestion that the X-ray luminosity of AXPs is due to mass accretion
from a fallback disk (Chatterjee et al. 2000; Alpar 2001) has triggered searches
for evidence of such disks. Searches for fallback disks around AXPs have
been conducted primarily in the optical and near IR bands. A realistic model for a putative
fallback disk should take into account the effects of irradiation by the neutron star's X-ray luminosity.
Irradiation is the
dominant source of disk luminosity for the outer disk at IR and longer
wavelengths. The radial temperature profile of a fallback disk has been computed by Perna et al. (2000)  and Hulleman et al. (2000) using the prescription given by \citet{Vrt90}. To estimate the temperatures, both
Hulleman et al. and Perna et al. assumed the same particular irradiation efficiency and found that their estimated optical flux values lie well above the values indicated by observations
of the AXPs 4U 0142+61 and 1E 2259+586. While Perna et al. suggested that the
difference might be due to a probable advection-dominated flow at the inner disk, Hulleman et al. drew
the conclusion that this model can only fit the data for an unrealistically small outer disk radius and that fallback-disk models were ruled out (see \S\ 4 for further discussion). Recent studies
show that irradiated-disk model fits to existing detections and upper limits
from AXPs allow the presence of accreting fallback disks with reasonable
irradiation parameters for all AXPs \citep{ErCa06} when the fits are not restricted to a particular
irradiation efficiency.

Recent results
by Wang et al. (2006, hereafter WCK06) from {\it Spitzer} data on the 
AXP 4U 0142+61 complement earlier data in the near-infrared and optical
(Hulleman et al. 2000, 2004, hereafter HvKK00 and HvKK04)  
and the detection of pulsed emission in the optical \citep{KM02}, thus
providing the first instance of a multiband data set that can be tested
against disk models. WCK06 interpret their data with a dust disk radiating
in the infrared, while the optical data are ascribed a magnetospheric
origin. They take the origin of the X-ray luminosity to be dissipation of magnetic energy in the neutron star's crust,
in accordance with the standard magnetar model.

Here we show that the entire unpulsed optical, near-IR, and far-IR spectrum can
be explained as due to a gaseous disk whose luminosity is provided by
viscous energy dissipation due to mass transfer through the disk, as well as
irradiation by the X-ray luminosity $L_{\mathrm{X}}$ from the neutron star.
Taking $L_{\mathrm{X}}$ to be an accretion luminosity, as we do, links the
irradiation contribution to the disk luminosity with the mass accretion rate $\dot{M}_{\mathrm{acc}}$
and, thereby, the mass inflow rate $\dot{M}$ through the disk and the viscous
dissipation contribution of viscous dissipation to the disk luminosity.
Luminosity in the optical band comes from the inner regions of the disk and
is supplied, to a significant extent, although not fully, by the local viscous energy dissipation in the inner disk. 
The innermost disk, which is dominated by dissipation, radiates mostly in the UV band but also contributes to the
total optical emission. Extension of the same irradiation model to outer regions of
the disk explains the IR radiation. The disk luminosity is dominated by the contribution of irradiation
at outer radii, where the thermal radiation peaks in the longer wavelength bands (see Table 1).

Pulsed emission in the optical \citep{KM02}
is due to magnetospheric emission powered by a disk-shell
dynamo in our model. It has been shown by \citet{CR91} that magnetospheric pulsar emission in optical and higher energy
bands can be sustained by an outer gap in the presence of a disk protruding into the magnetosphere. This disk model is different from the first disk-magnetosphere models \citep{michel81}. More recently, 
\citet{EC04} showed that the pulsed optical emission from 4U 0142+61 can be explained in terms of a
magnetospheric model with a disk and a neutron star surface dipole magnetic field of 
$B_{\mathrm{dipole}}\sim 10^{12}$-$10^{13}$ G. The magnetospheric emission is nearly 100\% pulsed in this
magnetosphere-disk model. The discovery of strongly pulsed optical emission \citep{KM02} has generally been perceived as excluding disk models, on the basis of the notion that magnetospheric emission cannot survive with a disk protruding into the light cylinder. The disk-magnetosphere model of  \citet{CR91} and the application by \citet{EC04} are important for fallback-disk models, since they demonstrate that a disk can actually be part of a magnetospheric mechanism that generates strongly pulsed emission.

The clustering of AXP and SGR rotational periods between 6 and 12 s was investigated by \citet{PM02}, who found that statistical clustering with a spin-down law of braking index $n\sim$ 2-4 implies that the final period of these systems must be about 12 s. This final period does not correspond to reaching the ``death line'' in the $P$-$\dot{P}$ plane which reflects the cut-off voltage for magnetospheric emission. Within isolated dipole-magnetar models, the alternative possibility for reaching a final clustering period, asymptotically, is by magnetic field decay on timescales commensurate with the AXP and SGR ages \citep{PM02}. An investigation of various magnetar field decay models (Colpi et al. 2000) has shown that having the X-ray luminosity data be associated with magnetic field decay is consistent with only one model, involving crustal Hall cascade, and for decay times less than about
$10^{4}$ yr. The presence of a fallback disk provides a natural explanation for period clustering of the AXPs \citep{chat}, as well as dim isolated (thermal) neutron stars (see, e.g., Haberl 2005) and SGRs \citep{Alpar01, EkA03}. Fallback-disk models do not address the origin of bursts, while magnetar models explain the bursts by employing the energy stored in very strong magnetic fields. The basic workings of the magnetar models require only local magnetic stresses and energies in the neutron star crust, and they can thus be sustained with higher multipole components of the field, while the period clustering and explanation of torques with a fallback disk require the dipole component of the magnetic field to be on the order of $10^{12}$-$10^{13}$ G. Hybrid models \citep{EkA03,ErA03} with $B_{\mathrm{multipole}}\sim10^{14}$-$10^{15}$ G and $B_{\mathrm{dipole}}\sim 10^{12}$-$10^{13}$ G are a possibility that incorporates a fallback disk.

In this paper, we evaluate the recent observations by WCK06 in
conjunction with older results (\S\ 2). We show that the data agree
quite well with a gaseous disk. In \S\ 3, we estimate the
characteristic lifetime of such a disk that has evolved through a
propeller phase and find that the disk's estimated age, 
depending on the torque model, is comparable to the $P_{\ast
}/2\dot{P}_{\ast }$ characteristic age estimate taken as if
dipole spin-down were effective. The discussion and conclusions are
given in \S\ 4.
 
\section{INFRARED AND OPTICAL RADIATION FROM THE DISK AROUND 4U 0142+61}

Recent {\it Spitzer} observations of the anomalous X-ray pulsar 4U 0142+61 have
detected the source in the 4.5 and 8 $\mu $m infrared bands (WCK06).
We will show that, taken together with the earlier observations in the optical (HvKK00, HvKK04; Morii et al. 2005),
these detections are consistent with the presence of a gaseous disk around
the neutron star. The optical detection must have a contribution from the
magnetosphere, as implied by the strongly pulsed part of the optical signal,
at 27\% amplitude in the R band, observed by
\citet{KM02}. \citet{EC04} showed that the strongly pulsed
emission from the magnetosphere can be accounted for either with a dipole-magnetar-based magnetosphere with surface dipole field strength $B_{\mathrm{%
dipole}}\sim 10^{14}$-$10^{15}$ G or with a magnetosphere with a disk in it, with $%
B_{\mathrm{dipole}}\sim 10^{12}$-$10^{13}$ G. The disk-magnetosphere model can provide
nearly 100\% pulsed magnetospheric emission in the optical. The presence of pulsed magnetospheric emission
does not rule out the presence of a disk protruding into the magnetosphere.
On the contrary, such a disk can be part of the magnetospheric circuit sustaining
pulsed emission, as shown by \citet{CR91}, and applied to 4U 0142+61
by \citet{EC04}. A pulsed component  is likely
to be present in the B, V, and IR bands. Uncertainties due to an unresolved
pulsed fraction that is constant across the bands cannot be distinguished from
uncertainties in other overall factors such as distance and disk inclination angle.
Possible variations in the relative pulsed fractions in different bands should be
taken into account in evaluating the model fits.
 Further, an important uncertainty arises from the interaction between inner disk and magnetosphere,
which might distort the inner disk's geometry and thus possibly lead to
time-dependent self-shielding at some inner disk regions. Observations of various AXPs 
in some of the optical/IR bands at different epochs show considerable same-source variations in relative amplitudes (see, e.g., Morii et al. 2005). These variations are not included in our
or other scientists' model fits, and indeed, it is not feasible to model such effects in a plausible
way. Keeping this in mind, we try to obtain the best fit to optical and IR data points.

Figure 1 shows the best model fit to all the data. Table 1 details the features of this fit. For each radiation band, the effective blackbody temperature representing that band and the radius $R$ in the disk corresponding to that effective temperature are listed. The last column gives the ratio of dissipation- to irradiation-powered disk flux at that radius. The total emission in each band is predominantly supplied by parts of the disk near the quoted radius, but of course, there are lesser contributions from all parts of the disk. In the best fit displayed in Figure 1 and Table 1, the representative radius, $1\times 10^10$ cm, that dominates the V-band emission draws about 30\% of its luminosity from viscous dissipation. The actual inner disk radius in this fit, $\rin = 10^9$ cm, is almost an order of magnitude smaller than the V-band representative radius. In the inner disk, the ratio of viscous dissipation to irradiation-supported flux is therefore higher. The radiation from the innermost disk is a major contributor to the UV radiation bands. The innermost disk, from $\rin$ to the V-band representative radius, also makes important contributions to the radiation in the optical bands. Therefore, the fits are quite sensitive to the value of the inner disk radius. The same
disk model accommodates the earlier optical and IR data (HvKK00, HvKK04; Morii et al. 2005) and
the {\it Spitzer} detections by WCK06. The value of $A_{V}$
indicated by the best model fit shown in Figure~1 is 3.5. This is within the
reasonable dereddening range $2.6<A_{V}<5.1$ for 4U 0142+61
(HvKK00, HvKK04) and coincides, fortuitously, with a recently derived estimate of $A_V=3.5 \pm 0.4$
(Durant \& van Kerkwijk 2006). For this model, all the model flux values remain within $\sim 30$\% of 
the data points. The error bars on the B-band data point give the 3 $\sigma$ upper limit (HvKK04).
Considering the uncertainties discussed above, the irradiated gaseous disk model is in reasonable
agreement with optical and IR data.

In the inner disk, viscous dissipation contributes significantly to
the disk luminosity, while in the outer and cooler parts of the disk, 
the luminosity is determined by X-ray irradiation from the neutron
star. Notably, the total disk luminosity of a gaseous disk,
including both irradiation and the intrinsic dissipation, does not
fall off in the optical range in the way expected for a dust disk as
employed by WCK06 in their Figure 3. Indeed, a gas disk model can
explain both the optical and the IR data, as shown in
Figure~1. The earlier conclusion by HvKK00 that an accretion disk is ruled out
 is no longer valid given the {\it Spitzer} observations. These
authors ruled out an accretion disk because their optical magnitude,
dereddened with $A_{V}=5.4$, does not fall on the disk curve
corresponding to $L_{\mathrm{X}}$ as an accretion luminosity,
with certain assumptions, a choice of $\dot{M}$, and a particular
irradiation reprocessing efficiency. Fitting the optical points to
the Rayleigh-Jeans end of a disk blackbody therefore
required, on the basis of the (only optical and near-IR) data
then existing, that any disk in 4U 0142+61 be truncated at
a certain outer radius $r_{\mathrm{out}}$. Now, with the new {\it Spitzer}
observations of WCK06, the disk is actually detected in the far-infrared. 
We can fit all the data in the optical and IR bands B,
V, R, I, J, H, K$_{s}$, 4.5 $\mu $m, and 8 $\mu $m (Fig.~1). A
gaseous disk emitting in this energy range extends from an inner
disk radius $r_{\mathrm{in}}\simeq 1\times 10^{9}$ cm to an outer
radius of about $10^{12}$ cm (see Table 1). The two mid-IR 
{\it Spitzer} data points put a lower limit on the outer disk radius
of around $10^{12}$ cm, but they do not constrain the extension of the
outer disk beyond $10^{12}$ cm. In our model, the outer disk radius is $r_{\mathrm{%
out}}=2\times 10^{12}$ cm. The quality of the fits does not change for even
larger values $r_{\mathrm{out}}$, since the contribution to the {\it Spitzer} bands at
4.5 and 8 $\mu$m from larger radii is not significant.

The model fits in the optical band are sensitive to the value of the disk's
inner radius. This is taken to be the Alfv\'{e}n radius,
\begin{equation}
r_{\mathrm{in}}\cong r_{\mathrm{A}}\cong f_{1}\dot{M}^{-2/7}(GM_{\ast
})^{-1/7}\mu _{\ast }^{4/7}, \label{rin}
\end{equation}%
where $M_{\ast }$ is the mass and $\mu _{\ast }$ is the dipole magnetic moment
of the neutron star. The factor $f_{1}\sim 0.5$-1 describes the uncertainty
in the precise location of the inner radius. The mass inflow rate $\dot{M}$
in the disk is larger than or equal to the mass accretion rate derived from
the X-ray luminosity,
\begin{equation}
\dot{M}_{\mathrm{acc}}\equiv \frac{L_{\mathrm{X}}R_{\ast }}{GM_{\ast }}\equiv f_{2}\dot{M}\leq \dot{M}, \label{accrate}
\end{equation}
where $R_{\ast }$ is the radius of the neutron star. As the star is spinning down while accreting, it must be in the propeller
regime but close to rotational equilibrium with the disk, according to the
fallback-disk model \citep{Alpar01}. Some of the mass inflow through the disk
may not be accreting under these circumstances. From the observed X-ray
luminosity of $L_{\mathrm{X}}\simeq 10^{35}$ ergs $\mathrm{s}^{-1}$, taking $R_{\ast }=10^{6}$ cm
and $M_{\ast }=1.4\,M_{\odot }$ we obtain $\dot{M}_{\mathrm{acc}}\simeq 5\times 10^{14}$ g $\mathrm{s}^{-1}$, which is related to the mass inflow rate $\Mdot$, depending on the choice of $f_{2}$. We first
report model fits with $\dot{M}=\dot{M}_{\mathrm{acc}}$ ($f_{2}=1$) and discuss
modifications of the results for $(f_{1},f_{2})<1$  in \S\ 4. For a given $\dot{M}$, the inner radius 
depends on the strength of the dipole magnetic moment of the neutron star.

We calculate the radial effective temperature profile $T_{\mathrm{eff}}(r)$
considering both the disk surface flux due to the intrinsic viscous
dissipation rate $D$ of the disk,
\begin{equation}
D=\frac{3}{8\pi }\frac{GM_{\ast }\dot{M}}{r^{3}}  \label{dissip}
\end{equation}%
(see, e.g., Frank et al. 2002) and the irradiation flux,
\begin{equation}
F_{\mathrm{irr}}=C\frac{\dot{M}_{\mathrm{acc}}c^{2}}{4\pi r^{2}}  \label{irflux}
\end{equation}%
\citep{SS73}, where the irradiation parameter $C$ includes the
effects of disk geometry, albedo of the disk surface, and the conversion
efficiency of the mass accretion into X-rays. In our calculations, we take $C
$ to be constant along the disk and leave its value as a free parameter. For
comparison with the data, we integrate the disk blackbody emission in each of
the optical and IR observational bands \citep[see][for details of calculations]{ErCa06}.
The model energy flux values presented in
Figures~1 and 2 were obtained with $C=1\times 10^{-4}$. While the irradiation
geometry of AXP disks might be somewhat different from those of low-mass X-ray binary (LMXB) 
disks (Ertan et al. 2006; Ertan \& \c{C}al{\i}\c{s}kan 2006), the values of $C$ that we find in our fits lie in the range 
expected for LMXBs ($\sim 10^{-4}$ to $10^{-3}$), in particular, based on disk stability analyses of  
soft X-ray transients (deJong et al. 1996; Dubus et al. 1999; Ertan \& Alpar 2002). With $C$'s in this range, the optical/IR observations for the other
three AXPs with available data can also be fitted with disk models with irradiation and viscous
dissipation \citep{ErCa06}.

In Figures~1 and 2, we present two different model fits. Figure~1 shows the
best fit, obtained with $A_{V}=3.5$, with the dipole field at the
neutron star's surface set to be $10^{12}$ G on the equator ($2\times 10^{12}$
G at the poles). After these fits were made, we learned of work on X-ray and
optical extinction and reddening (Durant \& van Kerkwijk 2006) reporting $A_{V}=3.5\pm 0.4$ for 4U 0142+61.
The inner disk radius $r_{\mathrm{in}}$ for the model in
Figure~1 is $10^{9}$ cm. The model fits are comparable for values of $r_{\mathrm{in}}$
up to a few times $10^{9}$ cm, with smaller $r_{\mathrm{in}}$
corresponding to larger $A_{V}$. The corotation radius for 4U
0142+61 is $r_{\mathrm{co}}\simeq 7\times 10^{8}$ cm, which is near the
inner disk radius we obtain from the best model fit. This implies that the
system is indeed near rotational equilibrium. The system is in the propeller
regime, as indicated by the spindown of the AXP. Nevertheless, most or all
of the inflowing disk matter is being accreted onto the neutron star rather
than being effectively propelled out of the system, on account of the
closeness to equilibrium. This is consistent with our model fits obtained
with the assumption that $f_{2}=1$.

The largest value of the dipole magnetic field allowed by the model fits 
corresponds to the lowest value of $A_{V}$. For the direction of  
4U 0142+61, reasonable values of the dereddening in the direction of
the AXP lie in the range $2.6<A_{V}<5.1$ (HvKK00, HvKK04).
Figure~2 shows the model fit for the largest value of the magnetic dipole
field. This was obtained by setting $A_{V}=2.6$, the minimum value in the
range considered, and allowing a 60\% discrepancy in the V band. Note that the discrepancy in V is $\sim 30$\% for the best fit, with $A_{V}=3.5$.  The inner radius of the disk is $r_{\mathrm{in}}=8\times 10^{9}$cm. The
corresponding dipole magnetic field value is found to be $2\times 10^{13}$ G
on the equator ($4\times 10^{13}$ G at the poles). The disk model fits
preclude higher values of the dipole component of the magnetic field. A stronger dipole field, $B_{\ast }>10^{14}$ G, would
cut the disk at much larger radii,above $10^{10}$ cm. This is not consistent
with the optical (R, V, B) data.

In comparing dereddened fluxes with models for various values of $B_{\mathrm{dipole}}$ and $A_{V}$, we find that the highest discrepancies always occur in the V band, where the dereddened fluxes are always somewhat larger than the model predictions. In the B band, the observations have yielded only an upper limit. Predictions in the B band from models that fit the V, R and IR bands are consistent with this upper limit.

\section{LIFETIME OF THE DISK AROUND 4U 0142+61}

Could a fallback disk remain active, exerting a torque on the neutron star, at the age of 4U 0142+61? In this section, we explore the evolution of the disk under various models for the torque between 
the disk and the neutron star. The characteristic lifetime of a fallback disk around a propeller
can be estimated as $\tau _{d}=M_{d}/\dot{M}$, where $\dot{M}$ is the mass
inflow rate in the disk.
For a steady configuration, the disk's mass-loss rate is $\dot{M}_{d}=-\dot{M}$. The mass-loss rate $\dot{M}_{d}$ will include the accretion rate
$\dot{M}_{\mathrm{acc}}$ onto the neutron star, as well as $\dot{M}_{\mathrm{out}}$, the mass
lost from the disk$-$neutron star system. In an effective propeller situation, $\dot{M}_{\mathrm{out}}$ may
make up most of $\dot{M}_{d}$. As the outer radius of the disk is not constrained by the data, the disk mass cannot be calculated from the fits. Instead, we shall assume plausible values, consistent with the observed disk to within an order of magnitude. We then use torque models to estimate the mass inflow rate $\Mdot$ through the disk from the observed spin-down rate
$\dot{\Omega}_{\ast }$.

The long-term spin history of a rapidly rotating
magnetized neutron star depends on the types of torques acting on it.
The net torque on a magnetized neutron star interacting with an accretion disk can
be expressed in general as
\begin{equation}
N_{\ast }\equiv I_{\ast }\dot{\Omega}_{\ast }=j\dot{M}(GM_{\ast }r_{\mathrm{%
in}})^{1/2},  \label{torque}
\end{equation}%
where $j$ is the nondimensional torque as a function of the fastness
parameter $\omega _{\ast }\equiv \Omega _{\ast }/\Omega _{\mathrm{K}}(r_{%
\mathrm{in}})$, $\Omega _{\ast }$ being the angular spin frequency of the
neutron star, $r_{\mathrm{in}}$ the inner disk radius, $\Omega _{\mathrm{K}%
}=(GM_{\ast }/r^{3})^{1/2}$ the Keplerian angular velocity, and $I_{\ast }$
the moment of inertia of the neutron star. Using the observed value of $\dot{%
\Omega}_{\ast }$ in equation (\ref{torque}), we estimate the mass inflow rate as
\begin{equation}
\dot{M}_{17}\simeq 0.4\,\omega _{\ast }^{-1/3}|j|^{-1}P_{\ast
}^{-7/3}I_{\ast 45}\left( \frac{\dot{P}_{\ast }}{10^{-12}\,\mathrm{s}\,%
\mathrm{s}^{-1}}\right) \left( \frac{M_{\ast }}{1.4\,M_{\odot }}\right)
^{-2/3},  \label{mdot}
\end{equation}%
where $\dot{M}_{17}$ is the mass inflow rate in units of $10^{17}$ g s$^{-1}$, $I_{\ast 45}$ is the neutron star's moment of inertia in units of $10^{45}$
g cm$^{2}$, and $P_{\ast }=2\pi /\Omega _{\ast }$ is the spin period of the
neutron star \citep{ErA04}.

In earlier epochs of its evolution, an AXP with a fallback disk is likely
to be in the propeller phase, in which the neutron star is rotating rapidly
with respect to the rotation rates in the inner boundary of the disk. Most
or all of the mass supplied by the disk is lost from the system rather than
being accreted by the propeller neutron star. This goes on until the neutron
star has spun down to rotation rates close to corotation with the disk. When
the star rotates only a little faster than the inner disk, a significant fraction
of the mass lost by the disk may be accreted onto the neutron star while, 
at the same time, the star is spinning down under torques applied by the disk.

The propeller-type spin-down torque first considered by \citet{IS75}
scales as $\left\vert N_{1}\right\vert \sim (\mu _{\ast
}^{2}/r_{\mathrm{A}}^{3})  \Omega _{\mathrm{K}}(r_{\mathrm{A}})/\Omega
_{\ast } $ \citep[see][]{ghosh95}. Depending on the spin rate of the
neutron star as compared with the sound speed at the magnetospheric boundary,
the scaling of the braking torques for subsonic and supersonic
propellers are given by $\left\vert N_{2}\right\vert \sim \mu _{\ast
}^{2}\Omega _{\ast }^{2}/GM_{\ast }$ and $\left\vert N_{3}\right\vert \sim
\mu _{\ast }^{2}\Omega _{\ast }\Omega _{\mathrm{K}}(r_{\mathrm{A}})/GM_{\ast
}$, respectively \citep{DP81,ghosh95}. In these models, the
magnetospheric radius is the Alfv\'{e}n radius (see eq. [1]). The spin-down torque
for a subsonic propeller, $N_{2} $, is independent of $\dot{M}$ and 
not used in the present work. The torques $N_{1}$ and $N_{3}$ can be
expressed as in equation (\ref{torque}) with $j=-\omega _{\ast }^{q}$ where
$q=-1$ and $q=1$ for $N_{1}$ and $N_{3}$, respectively. The braking torque
 considered by Wang \& Robertson ($N_4$; 1985) has the same form as $N_{3}$, but with a
particular estimate for the inner radius of the disk in the propeller stage,
that is, $N_{4}=-\omega_{\ast }\dot{M}(GM_{\ast }r_{\mathrm{in}})^{1/2}$, where
$r_{\mathrm{in}}=2^{-3/16}\Omega _{\ast }^{-3/8}\dot{M}^{-1/8}(2GM_{\ast })^{1/8}\mu _{\ast
}^{1/4}$. This scaling of the inner disk radius is simply that of the Alfv\'{e}n radius
when $r_{\mathrm{in}}=r_{\mathrm{A}}=r_{\mathrm{co}}\equiv
(GM_{\ast }/\Omega _{\ast }^{2})^{1/3}$, the corotation radius. All the
braking torques yield the same $\dot{M}$ when $\omega_{\ast }=1$,
in other words, when $r_{\mathrm{in}}=r_{\mathrm{co}}$.

We now estimate the lifetime of a fallback disk around 4U 0142+61
using these different types of propeller torque for disks of mass
$M_{d}=10^{-6}$ $M_{\odot }$ and $M_{d}=10^{-5}$ $M_{\odot }$. For each
torque model, solving equation (\ref{mdot}) with the implicit 
$\dot{M}$- and $B_*$-dependence of $\omega _{\ast }$ and $j(\omega _{\ast })$
gives the current value of the mass inflow rate $\dot{M}$. For
$M_{d}=10^{-6}\,M_{\odot }$, the inferred values for the lifetime of
the disk around 4U 0142+61 corresponding to the propeller torques
$N_{1}$, $N_{3}$, and $N_{4}$ can be written for $R_{\ast}=10^6$ cm and
$M_{\ast }=1.4\,M_{\odot }$ as $\tau _{1}\simeq 1.0\times
10^{5} B_{\ast 12}^{-4/9} \mathrm{yr}$, $\tau _{3}\simeq 4.8\times
10^{5} B_{\ast 12}^{8/3} \mathrm{yr}$, and $\tau _{4}\simeq
1.8\times 10^{5} B_{\ast 12}^{2/3} \mathrm{yr}$, respectively,
where $B_{\ast 12}$ is the surface dipole magnetic field
strength on the poles of the neutron star in units of $10^{12}$ G.
For $M_{d}=10^{-5}\,M_{\odot }$, all age estimates are greater by a
factor of 10. In the initial life of the neutron star, after a
brief initial accretion phase the propeller regime will start when
$r_{\mathrm{in}}>r_{\mathrm{co}}$ and will prevail until the inner
radius reaches the corotation radius again. As the inner radius of
the disk $r_{\mathrm{in}}$ approaches $r_{\mathrm{co}}$, the
propeller regime will allow accretion. For all torque models, the
condition $r_{\mathrm{in}}=r_{\mathrm{co}}$ yields the same mass
inflow rate,
\begin{equation}
\dot{M}=\Omega _{\ast }^{7/3}(GM_{\ast })^{-5/3}\mu _{\ast }^{2}
\end{equation}%
which yields for $M_{d}=10^{-6}\,M_{\odot }$ the critical age
estimate
\begin{equation}
\tau _{c}\simeq 4.67\times 10^{4}\;\mathrm{yr}\;B_{\ast 12}^{-2}.
\end{equation}%
As shown in Figure~3, $\tau _{c}$ is the minimum estimated lifetime for a  fallback
disk if the neutron star is spinning down under the action of propeller torques
without accreting any matter onto its surface. In our model, the AXPs experience propeller spindown (with some accretion) when $r_{\mathrm{in}}\geq r_{\mathrm{co}}$. The characteristic ages $\tau _{i}$ of the disk estimated  
from the propeller torques $N_{1}$, $N_{3}$, and $N_{4}$ are greater than $\tau _{c}$ when  
$r_{\mathrm{in}} > r_{\mathrm{co}}$, depending on the value of $B_*$. Figure 3 shows that the present age estimate
is larger than $\tau _{c}$ for all of the torques considered here if the
magnetic dipole field strength on the poles is greater than $6\times $ 10$^{11}$ G. If the end of the
propeller phase is taken to occur when $r_{\mathrm{in}}=r_{\mathrm{A}}=r_{\mathrm{co}}$, as
commonly assumed and adopted by WCK06, then $B_{\ast 12}\simeq 0.6$ is obtained
with $\tau _{c}\simeq 1.3\times 10^{5}~\mathrm{yr}=\tau _{1}=\tau_{3}=\tau _{4}$
for ${M}_{d}=10^{-6}\,M_{\odot }$ and $\tau _{c}\simeq 1.3\times 10^{6}~\mathrm{yr}=\tau _{1}=\tau_{3}=\tau _{4}$
for ${M}_{d}=10^{-5}\,M_{\odot }$ (see Fig.~3).

In scenarios with these propeller torques $N_{1}$, $N_{3}$, and $N_{4}$, 4U 0142+61
must have spent most of its lifetime in the propeller phase with little or no accretion, and it must
have entered its current phase of spin-down with accretion more recently. In the past, all or most
of the incoming disk matter would have been thrown out of the system as envisaged in the original 
propeller scenario \citep{IS75}. While in the present stage the gas disk is supplying the X-ray luminosity
of 4U 0142+61 by accretion, in the past the X-ray luminosity was not supplied not primarily
by accretion but by some other mechanism, such as dissipation of the magnetic field in the neutron star's surface, as in magnetar
models \citep{TD95}, or by internal energy dissipation in the neutron star, as expected
for neutron stars under torques but without initial cooling or accretion as a dominant source of surface thermal luminosity \citep{Alpar84, Alpar01}. 

If, on the other hand, there is accretion onto the neutron star over some
parts of its surface while a fraction of the infalling disk matter escapes
from the stellar magnetosphere as outflow or wind \citep[see][]{IK90},
the torque is given by
\begin{equation}
N_{5}= - \dot{M}_{\mathrm{out}}\Omega _{\ast }r_{\mathrm{in}}^{2}+\dot{M}_{%
\mathrm{acc}}\Omega
_{\mathrm{K}}(r_{\mathrm{in}})r_{\mathrm{in}}^{2}, \label{tfive}
\end{equation}
where $\dot{M}_{\mathrm{out}}=\dot{M}-\dot{M}_{\mathrm{acc}}$ by
conservation of mass, with the mass accretion rate,
$\dot{M}_{\mathrm{acc}}=L_{\mathrm{X}}R_{\ast }/GM_{\ast }$, expressed in terms of the X-ray luminosity $L_{\mathrm{X}}$.  We have defined
the mass outflow rate $\dot{M}_{\mathrm{out}}$ and the accretion
rate $\dot{M}_{\mathrm{acc}}$ as positive quantities. When the
negative propeller torque described by the first term dominates, the
net torque $N_{5}$ acts to spin down the neutron star. 
Such a spin-down phase allows one to explain the persistent-stage 
$L_{\mathrm{X}}$ of AXPs and SGRs with partial accretion from a
fallback disk, as we have done here. The first term in equation
(\ref{tfive}) is the braking torque on the neutron star due
to mass loss when $\dot{M}_{\mathrm{out}}<0$ \citep{ghosh95}. The
second term represents the angular momentum flux onto the neutron
star due to accreting disk matter. If accretion of matter onto the
neutron star has already started a switching off the effective
propeller mechanism within the recent history of 4U 0142+61,
we would expect to find commensurate estimates age and magnetic field
strength based on both $N_{5}$ and one of the propeller
torques $N_{1}$, $N_{3}$, and $N_{4}$, whichever is an apt
description of the actual torque effective in the system's past. In
Figure~3, we plot the age and polar surface magnetic field strength
estimates from $N_{1}$, $N_{3}$, $N_{4}$, and $N_{5}$. Note that the
spin-down torque $N_{5}$ and the propeller torque $N_{1}$ yield a
common estimate ($B_{\ast }\sim 6\times 10^{12}$ G)
for the dipolar component of the magnetic field. Even for the
smaller disk mass (${M}_{d}=10^{-6}\,M_{\odot }$), the torque models
$N_{3}$ and $N_{4}$ produce very large ages for $B_{\ast }>10^{12}$ G.
The torque $N_{5}$ represents the most plausible model, because it
incorporates accretion with spindown and yields reasonable ages, 
together with the torque $N_{1}$, for $B_{\mathrm{dipole}}\sim
10^{12}$-$10^{13}$ G.

Using the observed parameters of 4U 0142+61 and  torque $N_{5}$, the mass inflow
rate in the disk, for a given magnetic field strength, can be estimated
from equation (\ref{tfive}) along with the outflow rate $\dot{M}_{\mathrm{out}}$. We find $\dot{M}%
\simeq 1.9\times 10^{-11}\;M_{\odot }\;\mathrm{yr}^{-1}$ and $\dot{M}_{\mathrm{out}%
}\simeq 3\times 10^{-12}\;M_{\odot }\;\mathrm{yr}^{-1}$ for $B_{\ast
}=6\times 10^{12}$ G and $\dot{M}_{\mathrm{acc}}\simeq 1.6\times 10^{-11}\;M_{\odot }\;%
\mathrm{yr}^{-1}$ ($\sim 10^{15}$ g $\mathrm{s}^{-1}$). Note that
most of the mass lost by the disk accretes onto the neutron star.
This estimate implies $f_{2}=0.85$ with $f_{1}=1$.

The upper limit set on the lifetime of the disk by employing $N_{5}$
is much smaller  than the lifetime values estimated  using
the propeller torques $N_{3}$ and $N_{4}$. The spin-down age of the
pulsar estimated with the dipole spin-down formula, which of
course can only offer a rough comparison with the present model,
$P_{\ast }/2\dot{P}_{\ast}\sim 10^{5}$ yr, is comparable to $\tau
_{5}$, the inferred disk lifetime corresponding to the spin-down
torque $N_{5}$ in the accretion regime (see Fig.~3). This age
estimate is also consistent with those for the supernova
remnants associated with the AXPs 1E 2259+586 and 1E 1841$-$045. We
conclude that an active disk torquing down the star for an age of
$\sim 10^5$ yr is a tenable hypothesis, most likely with some
accretion for a significant part of its history up to the present.

With the assumption that $N_{5}$ is a good description of the
current torque, one can estimate the fraction $f_{2}$ of the mass
inflow that is accreted, for a given value of the disk inner radius,
without having prior knowledge of the mass inflow rate in the disk.
Equation (9) can be written in the form
\begin{equation}
\frac{\dot{M}_{\mathrm{out}}}{\dot{M}_{\mathrm{acc}}} =
(\frac{r_{\mathrm{co}}}{r_{\mathrm{in}}})^{3/2} + \frac{I_{\ast}
|\dot{\Omega}_{\ast}|}{\dot{M}_{\mathrm{acc}}\Omega_{\ast}r_{\mathrm{in}}^{2}},\label{tfivem}
\end{equation}
where we have substituted $N_{5}= I_{\ast}|\dot{\Omega}_{\ast}|$. Using the
observed spindown rate and the accretion rate inferred from the
luminosity (eq.[2]), we find $f_{2}$ = 0.6 for $r_{\mathrm{in}}$ =
10$^9$ cm and $f_{2}$ = 0.4 for $r_{\mathrm{in}}$ =
$r_{\mathrm{co}}$ = 7 $\times $ 10$^8$ cm.

\section{DISCUSSION AND CONCLUSIONS}

By combining earlier optical data (HvKK00, HvKK04; Morii et al. 2005) with  recent {\it Spitzer}
infrared data (WCK06), we have shown that the full data set is consistent
with the luminosity and spectrum expected from a gas disk. The optical
radiation is powered to a significant extent by viscous dissipation in the inner regions of the disk
while the IR radiation comes from outer regions where irradiation by the
central source is the primary source of the disk's local luminosity.

What is the maximum dipole magnetic moment $\mu _{\ast }$ that is compatible with
these disk model fits to the data? From equations (1) and (2),
we obtain
\begin{equation}
\mu _{\ast }=f_{1}^{-7/4}f_{2}^{-1/2}r_{\mathrm{in}}^{7/4}\dot{M}_{\mathrm{acc%
}}^{1/2}(GM_{\ast })^{1/4}.
\end{equation}
Using $f_{1}=0.5$ and $f_{2}=1.0$ and employing the maximum
$r_{\mathrm{in}}\cong 8 \times 10^{9}$ cm that is compatible with our fits,
for a minimum $A_{V}=2.6$ we find the maximum possible $\mu _{\ast }$
to be $5.8\times 10^{31}$ G cm$^3$ corresponding to a maximum
surface dipole field of $1.2 \times 10^{14}$
G on the poles of a neutron star of mass $1.4\,M_{\odot }$ 
and radius $R_{\ast}=10^{6}$ cm. A further increase in the upper
limit for the magnetic field can be obtained with smaller values of $f_{2}$. The minimum $f_{2}$ that allows a fit within 60\%  V-band deviation is around 0.3,  together with $r_{\mathrm{in}}\simeq 1.0
\times 10^{10}$ cm. For these extremes of parameter values, we
obtain $B_{\ast}\simeq 3 \times 10^{14}$ G on the pole of the
neutron star. These estimates should be noted together with the
uncertainties due to disk-magnetosphere interaction, possible
shielding effects, and time-dependent variations as discussed in
\S\ 2. 

We note that if the disk inner radius is large,
$r_{\mathrm{in}} \gg r_{\mathrm{co}}$, the neutron star is a fast
rotator, far from rotational equilibrium with the disk, and likely
to be in the strong propeller regime. If the simple description of
propeller spin-down given by equation (9) is roughly correct, as is
likely the case, the large moment arm provided by a large
$r_{\mathrm{in}}$, together with mass outflows at least comparable
to (and probably larger than) the mass accretion rate onto the star
will yield spindown rates much larger than that observed:
\begin{equation}
|\dot{\Omega}_{\ast}| \sim
\dot{M}_{\mathrm{acc}}\Omega_{\ast}r_{\mathrm{in}}^{2}.
\end{equation}
A disk with $r_{\mathrm{in}}\simeq 1.0 \times 10^{10}$ cm, which 
allows for a magnetic dipole as large as $B_{\ast}\simeq$ 3 $\times
10^{14}$ G, requires a spindown rate of $|\dot{\Omega}_{\ast}|
\sim $ 10$^{-10}$ rad s$^{-2}$. This is much larger than the
observed $|\dot{\Omega}_{\ast}| \simeq $ 1.7 $\times$
10$^{-13}$ rad s$^{-2}$. Here we have a strong indication against 
values of $r_{\mathrm{in}}$ large enough accommodate dipole magnetic
fields in the magnetar range.

As our disk models depend only on the mass inflow rate
$\dot{M}$ inside the disk and the irradiation parameter $C$, it could
be argued that the entire mass inflow actually turns into an
outflow and none of the mass flowing through the disk is accreted
onto the neutron star, the X-ray luminosity being due to some energy
release mechanism other than accretion. The observation that 
the inner disk radius in our best fit is so close to the co-rotation
radius argues against this. Indeed our fits yield the inner radius required by the optical
observations. The actual inner radius may be even smaller and closer
to $r_{\mathrm{co}}$, but we do not have the possibility of
observing such a highly absorbed source in the UV band to check.
The rather small value of the spin-down rate compared with what one
would expect with a substantial mass outflow from a larger inner
radius also indicates that this is a weak propeller, close to
equilibrium. It is therefore likely that there is mass accretion.

Could the disk of our model fits be passive, with no viscous
dissipation, no mass inflow, and no accretion at all onto the neutron
star? Under this hypothesis, the disk luminosity would be generated
entirely by irradiation by the neutron star's X-ray luminosity,
which in turn is not due to accretion. It is unlikely 
that there is no accretion at all, 
if there is a gas disk with the properties indicated by our model fits, 
which are sensitive to the inner disk radius and require
temperatures above $ 10^{5}$ K in the inner disk. The disk cannot be
neutral and passive at these temperatures; it has to be ionized,
whether it is pure hydrogen or contains heavy elements. An ionized
disk would have viscosity generated by the magnetorotational
instability \citep{BH91}, which will result in mass inflow through
the disk and interaction between the disk and the neutron star
magnetosphere, as well as some mass accretion onto the neutron star,
depending on the closeness to equilibrium. Ionization temperatures
for hydrogen disks are of the order of $6000$ K. For fallback disks
with heavy metals (Werner et al. 2007), the ionization temperatures could be as
low as $\sim 2500$ K. In hydrogen disk simulations of soft X-ray
transients and dwarf novae, two different viscosity parameters are
employed above and below the ionization temperatures to account for
the observations of these systems. In these models, the viscosity
parameter decreases by a factor of 0.01-0.5 as the
temperatures decrease to below the ionization temperature,
$T_{\m{i}}$.  This indicates that even the disk regions with $T <
T_{\m{i}}$ remain active and that the critical temperature below which
the entire disk enters a passive state is likely to be much
lower than the ionization temperatures.
Early studies by Gammie (1996) on protoplanetary disks suggested the possibility that a weakly ionized disk might become passive in its dense regions in the mid-plane while accretion proceeds over its surface. Recent work by Inutsuka $\&$ Sano (2005), however, indicates the existence of various physical mechanisms that can provide, sustain, and homogenize a sufficient degree of ionization for the magnetorotational instability (MRI) to work through most regions of a protoplanetary disk even at temperatures as low as 300 K. According to Inutsuka $\&$ Sano, once MHD turbulence starts to prevail in a disk, it seems quite difficult to switch it off.  
The state of turbulent viscosity keeps the temperature and other parameters such that the ionization level required for MRI is self-sustained.

The presence of the disk imposes constraints only on the dipole component of the magnetic field. 
The very strong magnetic fields that are employed to explain SGR and AXP bursts in the magnetar model may well 
reside in the higher multipole structure of the neutron star's surface magnetic field. In this hybrid situation, 
dissipation of the magnetar fields' energy in the neutron star crust may  contribute to the surface luminosity along with accretion. In this case,  $\dot{M}_{\mathrm{acc}}$ will have a value less than that inferred by taking the X-ray luminosity to be fully due to accretion, as in equation (2).
 
The object 4U 0142+61 is the anomalous X-ray pulsar with the most detailed
optical and IR observations. It would  be of great interest to check and
extend the current database with further 
observations, in particular with {\it Spitzer}. It is important to realize that magnetar fields in higher multipoles and a dipole field with  $B_* \la 10^{13}$ G, to accomodate a disk with the equilibrium period  range of observed period clustering, can coexist in the neutron star. The
search for disks in other AXPs has also provided some data in
the optical and near-infrared that are consistent with gas disk
models, as discussed in \citet{ErCa06}. Time-dependent variations
in the X-ray luminosity could be smeared out during the X-ray
reprocessing at the outer disk. On the other hand, variations in the
mass inflow rate modify the disk emission in the optical bands
before the changes have been observed in X-rays, because of varying
accretion rates. Any observed signatures in the optical bands
followed by similar variations in X-rays with viscous time delays
(minutes to days) would be a clear indication of ongoing disk
accretion onto the neutron star.  At present, 4U 0142+61 seems to
be the best source with which to test this idea.

Finally, we comment on the passive dust-disk model proposed by
Wang et al. (2006). Their detection of 4U 0142+61 in the mid-infrared has
firmly indicated, for the first time, the presence of a disk around
an isolated neutron star. They fitted the data with a
two-component model, using a power law to fit the optical and near
IR data, which they ascribe to magnetospheric emission. This leaves
only their recent discovery, the {\it Spitzer} detections in the mid-IR
band, to be explained by a disk. Hence, while the evidence in that
band clearly points to a disk, the association of only this narrow-band 
emission with the disk constrains it to be a dust disk beyond
the magnetosphere. Questions arising from this model concern why
this disk did not generate some viscosity to extend into an active
gaseous disk, how it cooled and remained cold and confined, and
whether it is stable against the radiation pressure of the pulsar.
With an inner disk radius $r_{\m{in}}\sim 2 r_{\m{lc}}$, where
$r_{\m{lc}}$ is the light cylinder's radius, such a passive dust disk
could be stable against the radiation pressure of the magnetospheric
pulsar emission (Ek\c{s}i \& Alpar 2005). This stability is
valid for $r_{\m{lc}} ~\la ~r_{\m{in}}~\la ~2 r_{\m{lc}}$ for orthogonal rotators
and for even wider range for non-orthogonal rotators according to the investigation of
Ek\c{s}i \& Alpar (2005), the stability issue should be kept in mind
for dust disks beyond the light cylinder.

\acknowledgements

We acknowledge research support from the Sabanc\i\ University Astrophysics and Space Forum.
M. A. A. acknowledges support from the Turkish Academy of Sciences.

\clearpage

\begin{table}

\begin{center}
\caption{Model Temperatures and Radii Corresponding to Different Optical and Infrared Bands.}

\vspace{1 cm}
\begin{tabular}{l|c||c||c}
\tableline\tableline
Band& $T_{\m{BB}}$ (K)& $R$ (cm)& $D/\Firr$
\\
\hline
$B$& 6516 & $7.0\times 10^9$& 0.45
\\
\hline
V& 5263&$ 1.0\times 10^{10}$
&0.3
\\
\hline
R& 4454& $1.4\times 10^{10}$&0.30
\\
\hline
I & 3585& $2.2\times 10^{10}$&$ 0.14$
\\
\hline
J& 2377& $4.8\times 10^{10}$&0.07
\\
\hline
H&1779&$7.9\times 10^{10}$&0.04
\\
\hline
K$_{\m{s}}$& 1324&$1.4\times 10^{11}$&0.02
\\
\hline
4.5 $\mu$m & 644&$5.9\times 10^{11}$&0.005
\\
\hline
8 $\mu$m & 362&$1.9\times 10^{12}$&0.002
\\
\tableline

\end{tabular}
\end{center}
\tablecomments{The rightmost column shows the ratio of viscous dissipation rate to irradiaton flux}
\end{table}

\clearpage
\begin{figure}
\vspace{-10 cm}
\hspace{-4 cm}
\includegraphics{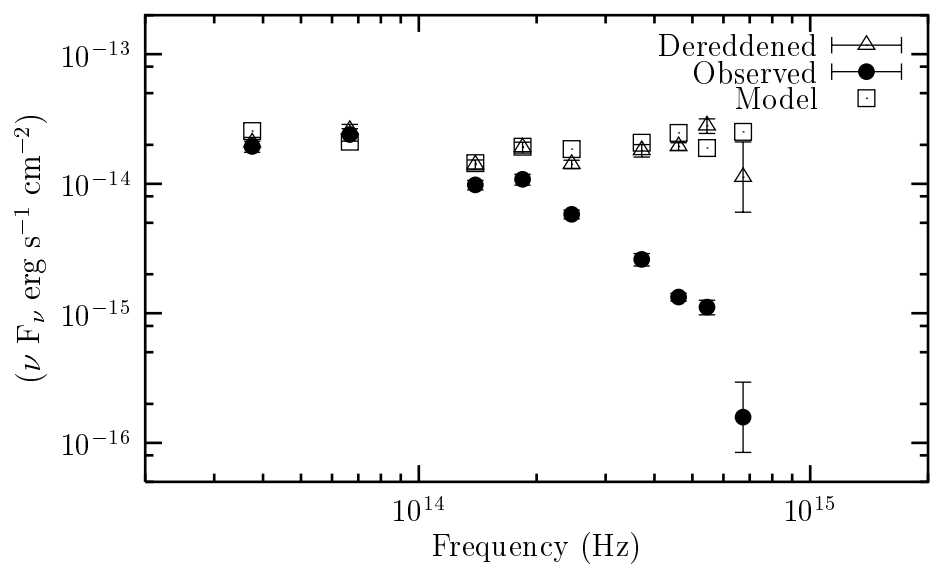}
\vspace{-11 cm}
\caption{Energy flux data and irradiated disk model values for
4U 0142+61 in the optical and infrared bands (B, V, R, I, J, H, K$_s$,
4.5 $\mu$m and 8 $\mu$m). Circles are the observed (absorbed) data (taken from HvKK00 [V, R, I], HvKK04 [B, K$_s$], Morii et al. 2005 [H, J], and WCK 2006 [4.5 $\mu$m and 8 $\mu$m]), and triangles are data dereddened with $A_V=3.5$. Squares are the irradiated-disk model energy flux values (see \S\ 2).}
\end{figure}

\clearpage
\begin{figure}
\vspace{-10 cm}
\hspace{-4 cm}
\includegraphics{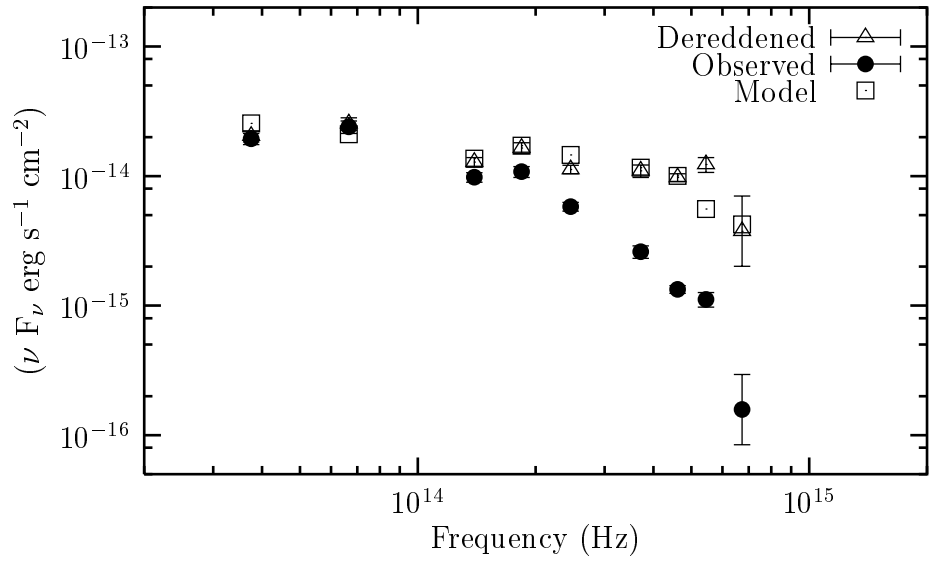}
\vspace{-11 cm}
\caption{Same as Fig. 1, but with dereddennig of  $A_V=2.6$ (triangles).}
\end{figure}

\clearpage
\vspace{-2 cm}
\begin{figure}
\includegraphics{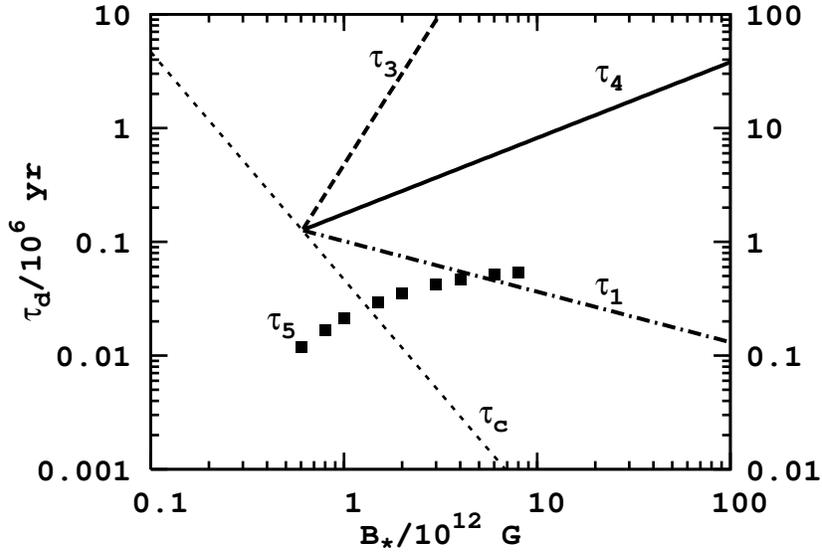}
\vspace{1 cm}
\caption{Disk lifetime as a function of stellar magnetic field strength
for different propeller torques. Squares show the disk lifetime values
estimated from the spin-down torque $N_{5}$ in the accretion
regime. The left vertical axis shows the estimated lifetime range for a disk of mass $M_{d}=10^{-6}\,M_{\odot }$.
The vertical axis at right is for $M_{d}=10^{-5}\,M_{\odot }$.}
\end{figure}

\end{document}